%% file: Theory_nonkramers1106.tex
\documentclass[english,aps,prb,reprint,superscriptaddress]{revtex4-2}
\usepackage[T1]{fontenc}
\usepackage{color}
\usepackage{babel}
\usepackage{amsmath}
\usepackage{amssymb}
\usepackage{graphicx}
\PassOptionsToPackage{normalem}{ulem}
\usepackage{ulem}
\usepackage[unicode=true,pdfusetitle,
 bookmarks=true,bookmarksnumbered=false,bookmarksopen=false,
 breaklinks=false,pdfborder={0 0 1},backref=false,colorlinks=true]
 {hyperref}
\hypersetup{
 allcolors=blue}

\makeatletter

\providecommand{\tabularnewline}{\\}
\usepackage{txfonts}
\newcommand{\hide}[1]{}

\makeatother

\begin{document}
\title{Theory of magnetism for rare-earth magnets on 
	the Shastry-Sutherland lattice with non-Kramers ions}

\author{Guijing Duan}
\affiliation{School of Physics and Beijing Key Laboratory of Opto-electronic 
Functional Materials \& Micro-nano Devices, Renmin University of China, Beijing 
100872, China}
\author{Rong Yu}
\email{rong.yu@ruc.edu.cn}
\affiliation{School of Physics and Beijing Key Laboratory of Opto-electronic 
Functional Materials \& Micro-nano Devices, Renmin University of China, Beijing 
100872, China}
\affiliation{Key Laboratory of Quantum State Construction and Manipulation 
(Ministry of Education), Renmin University of China, Beijing, 100872, China}
\author{Changle Liu}
\email{liuchangle89@gmail.com}
\affiliation{School of Engineering, Dali University, Dali, Yunnan 671003, China}

\begin{abstract}
Motivated by the rapid experimental progress on the rare-earth 
Shastry-Sutherland lattice magnets, we propose 
a generic effective spin model that describes interacting  
non-Kramers local moments 
on the Shastry-Sutherland lattice.  
We point out that the
local moments consist of both magnetic dipole and quadrupole components 
and the effective model turns out to be an extended XYZ model with an intrinsic
field that accounts for the crystal field splitting. We then study 
the ground-state phase diagram of the model and 
find 
that 
pure quadrupole orders, which are 
invisible  
to conventional experimental probes, can be stabilized over a broad regime. In 
particular, we 
show that a hidden ``1/3 magnetization plateau'' with quadrupole orders 
generally exists 
and discuss its
experimental signatures.  
Finally, we discuss
the relevance of our results to the rare-earth Shastry-Sutherland lattice 
magnets Pr$_{2}$Ga$_{2}$BeO$_{7}$ and Pr$_{2}$Be$_{2}$GeO$_{7}$.
\end{abstract}
\date{\today}
\maketitle

\section{Introduction}


Geometrically frustrated quantum magnets 
emerge as a playground
to study various  
unconventional quantum phenomena characterized
by suppressed long-range magnetic order and peculiar low-energy dynamical
behaviors \cite{lacroix2011introduction,savary2016quantum,balents2010spin}.
One prominent example is 
the Shastry-Sutherland lattice (SSL) 
magnets, 
which host a variety of novel phases owing
to the strongly frustrated interactions between localized magnetic moments. 
This system consists of orthogonal spin dimers with frustrated inter-dimer 
interactions on a two-dimensional (2D) lattice
\cite{shastry1981exact}. 
Experimentally, it has been widely recognized that the low-temperature phases 
of the quantum magnet SrCu$_{2}$(BO$_{3}$)$_{2}$ can be well described by the 
$S=1/2$ AFM Heisenberg model on the SSL.
In this compound, the competing interactions result in a 
dimer singlets (DS) 
ground state where 
spin dimers along the diagonal bonds form singlets
\cite{miyahara1999exact}. By applying a hydrostatic pressure, the
system undergoes a phase transition to a plaquette valence bond solid (PVBS) 
and then to a N\'eel antiferromagnetic (AFM) ordered
state \cite{guo2020quantum,haravifard2016crystallization}.
The presence of PBVS-to-AFM phase transition makes this system
an ideal experimental platform to investigate the exotic physics of deconfined 
quantum criticality (DQC) \cite{cui2023proximate,zayed20174,lee2019signatures}.
Moreover, the ground state can also be tuned by applying an external magnetic 
field, where various fractional magnetization plateaus emerge, reflecting 
the strong geometrical frustration in this system 
\cite{dorier2008theory,momoi2000magnetization,nomura2023unveiling}.


Recently, research interest on quantum magnetism has been extended to 
systems with strong spin-orbit coupling (SOC)
\cite{zhou2017quantum,broholm2020quantum,takagi2019concept,trebst2022kitaev,KITAEV06,hallas2018experimental,bramwell2001spin,li2020spin,
liu2018selective,liu2020intrinsic,bramwell2001spin,liu2018selective,shen2019intertwined,liu2020intrinsic,liu2020intrinsic,savary2015probing,liu2018selective,shen2019intertwined}.
	In these systems, the
effective moments are formed by a combination of spin and orbital angular
momenta, and the interactions among the moments are usually
anisotropic in both real space and spin space.
This exchange anisotropy can induce dynamic frustration, resulting 
highly non-trivial quantum phases of matter such as quantum spin liquids (QSL) 
\cite{balents2010spin,savary2016quantum,zhou2017quantum,broholm2020quantum}, 
where local moments remain highly correlated while strongly fluctuating. 
Following such spirit, there have been quite intense experimental and 
theoretical activities on exploration of relevant systems, such as 
honeycomb Kitaev 
magnets \cite{KITAEV06,takagi2019concept,trebst2022kitaev},
 pyrochlore spin 
ice \cite{hallas2018experimental,bramwell2001spin}, and 
triangular lattice materials 
\cite{li2020spin}.
Moreover, in strong SOC magnets
the moments may contain 
high-order multipolar  
components, such as quadrupoles and octupoles  
\cite{liu2018selective,shen2019intertwined,liu2020intrinsic}. This can give rise to   
a rich variety of novel phases and 
quantum critical behaviors. However, multipolar moments and their ordering 
cannot be directly probed by traditional experimental techniques, such as  
neutron diffraction. This puts
a crucial challenge on how to detect 
these hidden phases experimentally \cite{savary2015probing,liu2018selective,shen2019intertwined,liu2020intrinsic}. 

To gain a deeper understanding on the experimental behaviors of such 
strong SOC materials, it is essential to construct microscopic models that are 
able to properly describe the interacting local moments. 
Microscopic theories on Kitaev, 
pyrochlore and triangular lattice materials have been well developed \cite{liu2018selective,liu2020intrinsic,bramwell2001spin}. However, 
existing theory for magnets on the SSL 
are still limited 
to strongly Ising anisotropic
systems (TmB$_{4}$, \emph{etc.}) \cite{song2020abnormal,
	lancon2020evolution,nagl2024excitation,ashtar2020new,verkholyak2022fractional,
	regeciova2023magnetocaloric,muto201211b,ishii2021magnetic,ishii2020high}.
A generic model describing the full source of quantum
fluctuations in SSL systems is highly demanded given the growing experimental reports on relevant materials recently \cite{ashtar2021structure,brassington2024magnetic,brassington2024synthesis, 
yadav2024observation,li2024spinons,liu2024distinct,ashtar2021structure}.


The major difficulty in constructing a generic theory of strong 
spin-orbit-coupled 
SSL magnets comes from the low crystal symmetry of 
these systems. As a comparison, for the extensively studied 
triangular, honeycomb and pyrochlore lattice
systems
, the point group symmetry 
at the magnetic
ion contains at least a three-fold rotation. 
The rotational symmetry puts strong restrictions on the form of crystal
field wave functions, and constrains the magnetic moment directions
along or perpendicular to the three-fold axis. Moreover, they also
strongly restrict the form of interactions and the number of model parameters, 
making the effective Hamiltonian
of relatively simple and convenient for numerical studies. On the other hand, 
for Shastry-Sutherland
magnets,
the symmetry of crystal field environment around
rare-earth ions is 
much lower. For example, RE$_{2}$Ga$_{2}$BeO$_{7}$
and RE$_{2}$Be$_{2}$XO$_{7}$ (X=Si, Ge) \cite{ashtar2021structure,li2024spinons,liu2024distinct,brassington2024synthesis}
the local environment of RE$^{3+}$ only contains no rotational symmetry but 
only a mirror reflection (as marked in Fig. \ref{fig:crystal}).
As a consequence of the low symmetry, the directions of the local moments
are no longer fixed to be parallel or perpendicular to the crystal
plane, but in general form a certain angle with respect to the crystal
plane, where the angle is determined by the detail of crystal field wave
functions. Moreover, the multi-sublattice structure of the SSL makes the 
angle sublattice dependent, which 
further complicates the analysis of experimental data and the modeling.


In this paper, we develop a generic effective spin model for rare-earth 
Shastry-Sutherland magnets. For clarity, in this work we only  
consider the case of non-Kramers magnetic ions, and the Kramers case will be 
discussed in a separate work \cite{Duan2024}. For
Kramers ions, the crystal field levels always form degenerate
Kramers doublets protected by the time-reversal symmetry. However, for 
non-Kramers ions there is no symmetry to protect any crystal field degeneracy, 
namely, all crystal field levels are in principal non-degenerate singlets. The 
presence of crystal field splitting generates an intrinsic field for the local
moments. In addition, for a non-Kramers moment only one
component (denoted as longitudinal) behaves as a magnetic dipole, while the 
other two components (transverse ones)
are even under the time reversal symmetry and behave as quadrupoles. Based on 
symmetry analysis, we find that the effective spin model  
takes the form of 
an extended XYZ model supplemented by an intrinsic field that 
originating from the crystal field splitting. To reveal the physical roles of 
the XYZ exchange anisotropy and the intrinsic field respectively, we first
switch off the intrinsic field in the model and discuss the phase diagram in 
the presence of XXZ- and XYZ-type anisotropy. Then we examine the influence of 
the intrinsic field to the XYZ anisotropic model. Given that the effective spin
model has too many model parameters, obtaining a full phase
diagram is impractical. To manage computational complexity, we have
selected a few representative parameter points and explore the implications
of the intrinsic field, respectively. This approach allows us to gain
insights into the underlying physics without the need for exhaustive
computational resources.

The rest of the paper is organized as follows. In Sec. \ref{sec:Model},
we provide a symmetry analysis on the structure of local moments and derive the 
effective Hamiltonian. In Sec. \ref{sec:BField}
we explore how the system couples to the magnetic field. 
As a result of the sublattice dependent local moment environment, this coupling 
is quite different from ordinary spin systems, leading to a complicated 
magnetic response to the applied field. We discuss how to extract information 
about the exchange couplings from magnetic susceptibility measurements.
In Sec. \ref{sec:Phase diagram} we present the numerical phase diagram
demonstrating the implications of XXZ- and XYZ-type exchange anisotropy
as well as the intrinsic field, respectively. Due to the quadrupole
nature of the transverse spin components, several quadrupolar ordered states 
can be stabilized in the phase diagram. 
Although the quadrupolar order parameters are not directly observable in most 
conventional probes, 
the underlying structures of these hidden orders 
are reflected in the dynamical excitation spectra, which are visible via 
neutron and light scattering measurements. In Sec. \ref{sec:Discussions} we 
summarize our
results and discuss the relevance of our work to a number of rare-earth SSL 
magnets recently reported, including
Pr$_{2}$Be$_{2}$GeO$_{7}$ and Pr$_{2}$Ga$_{2}$BeO$_{7}$.

\section{Generic effective spin model\label{sec:Model}}
\begin{figure}
	\includegraphics[width=1\linewidth]{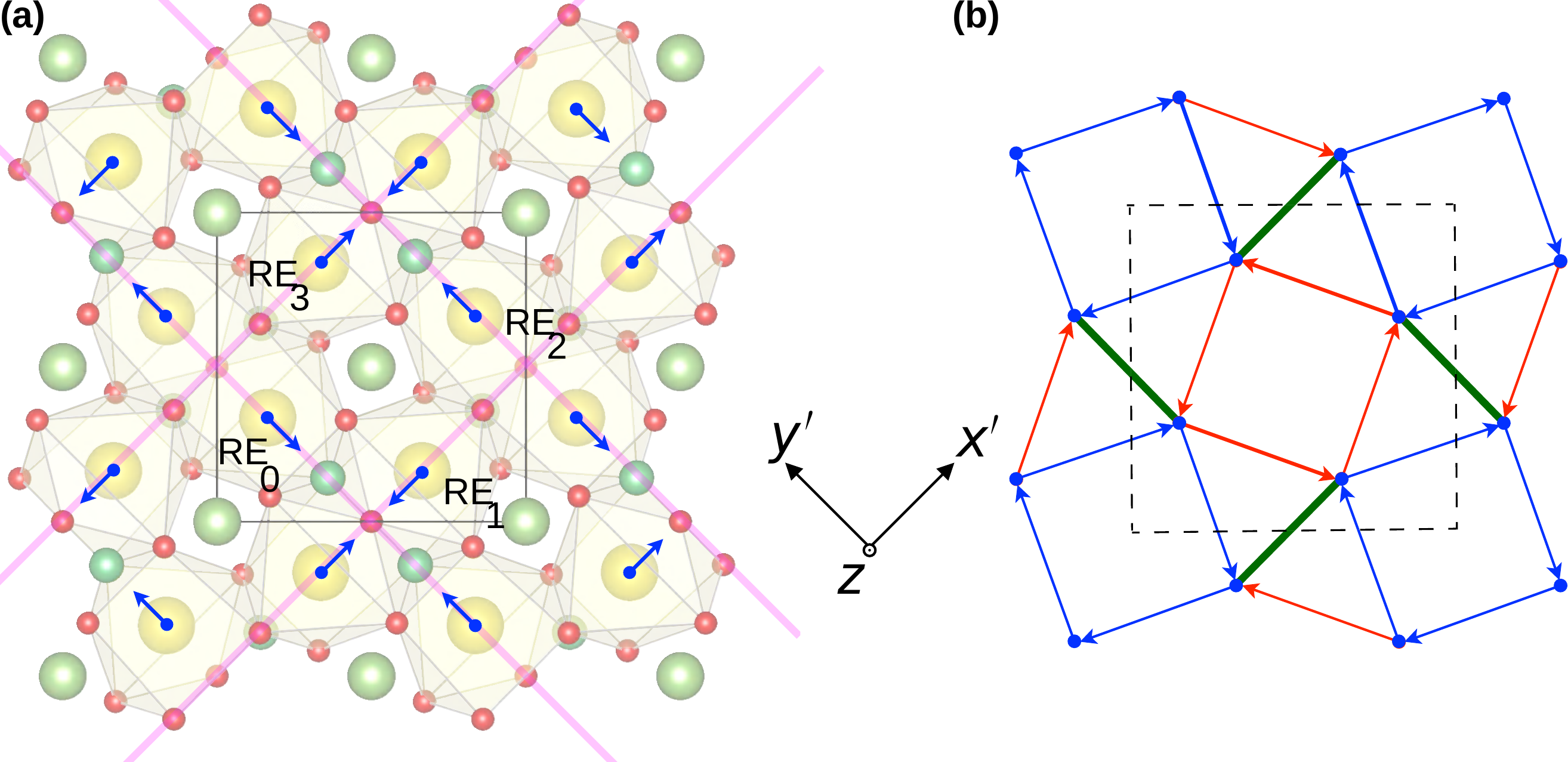}
	\caption{Crystal structure of RE$_{2}$Ga$_{2}$BeO$_{7}$ and sketch of 
		the SSL. (a). RE$^{3+}$ and
		O$^{2-}$ are denoted by yellow and red balls, respectively. Mirror
		planes of RE$^{3+}$ are indicated by purple lines. The in-plane 
		direction
		of the local dipole axes $\mathbf{n}_{i}$ are labeled by blue arrows. 
		(b). The intra-dimer bonds of the SSL are indicated by thick dark green 
		lines. The 
		$\eta_{ij}=\pm1$
		inter-dimer bonds in Eq. (\ref{eq:inter-dimer}) are marked
		as red and blue, respectively, with the directions $i\rightarrow j$ are 
		marked by arrows.}
	\label{fig:crystal}
\end{figure}
\hide{
\begin{figure}[t!]
	\includegraphics[width=1\columnwidth]{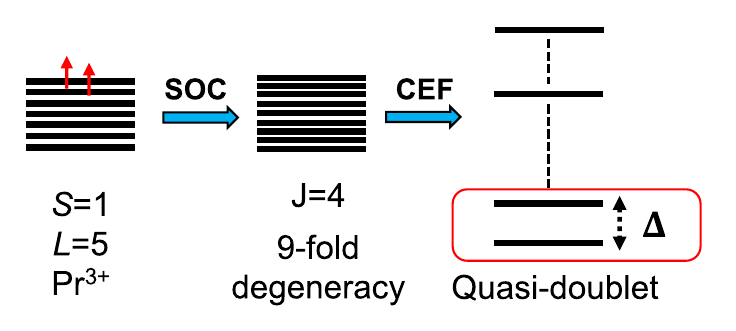}
	
	\caption{Crystal field scheme of Pr$^{3+}$ in the rare-earth SSL magnet 
		Pr$_{2}$Be$_{2}$GeO$_{7}$. The red arrows refer to Pr $4f$ electrons. 
		The SOC 
		couples orbital and spin angular momenta $L$ and $S$ to the total 
		angular 
		momentum $J=4$. Under crystal electric field (CEF) the 9-fold multiplet 
		corresponding to $J=4$ further splitts to 9 non-degenerate singlets. 
		The two 
		lowest crystal levels (enclosed by the red box) are close in energy and 
		form a 
		quasi-doublet which is well separated in energy to other crystal field 
		levels.  
		This quasi-doublet accounts for the low-temperature magnetism of the 
		material.}
	
	\label{fig:CEF}
\end{figure}
}
\subsection{Crystal field scheme}

For definity, we consider the crystal structure of SSL compounds 
RE$_{2}$Be$_{2}$GeO$_{7}$,
where RE$^{3+}$ refers to non-Kramers ions, RE=Pr, Sm, Tb, Ho, Tm. 
Here we take 
RE=Pr
as an example to illustrate the crystal field scheme (shown in Fig. 
\ref{fig:CEF}). 
The non-Kramers Pr$^{3+}$ ion has total orbital
angular momentum $L=5$ and total spin angular momentum $S=1$. The
strong spin-orbit coupling couples $L$ and $S$ to the total angular
momentum $\mathcal{J}=4$. In a crystal field environment, these nine-fold 
multiplets
further split into various crystal field levels. Since the point group
symmetry of the Pr$^{3+}$ site contains only a mirror reflection
(mirror planes shown by purple lines in Fig. \ref{fig:CEF}), all
crystal field levels form non-degenerate singlets \cite{liu2024distinct}.
If the crystal field gap were sufficiently large, 
the low-energy property of the system would be dominant by the lowest singlet 
with effective spin zero, and hence we would not expect any magnetism. However, 
in experiments the system become magnetized under external
magnetic field, and in thermodynamic measurements the magnetic entropy
saturates to $\sim0.89R\ln2$ at around 15 K \cite{liu2024distinct}. These 
strongly
implies 
that the magnetism is contributed by a quasi-doublet
CEF ground states, which are  
well separated in energy to other excited crystal field
levels, as illustrated in Fig. \ref{fig:CEF}.
\begin{figure}[t!]
	\includegraphics[width=1\columnwidth]{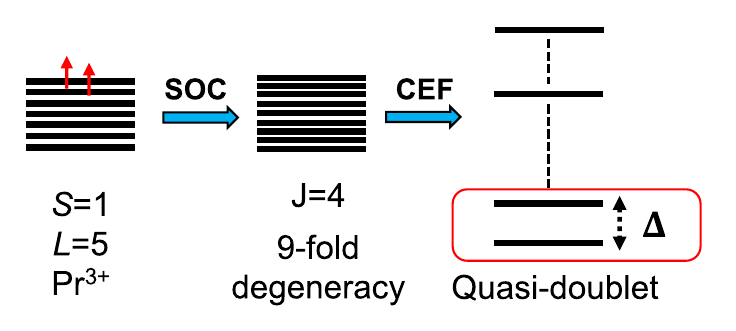}
	
	\caption{Crystal field scheme of Pr$^{3+}$ in the rare-earth SSL magnet 
		Pr$_{2}$Be$_{2}$GeO$_{7}$. The red arrows refer to Pr $4f$ electrons. 
		The SOC 
		couples orbital and spin angular momenta $L$ and $S$ to the total 
		angular 
		momentum $J=4$. Under crystal electric field (CEF) the 9-fold multiplet 
		corresponding to $J=4$ further splitts to 9 non-degenerate singlets. 
		The two 
		lowest crystal levels (enclosed by the red box) are close in energy and 
		form a 
		quasi-doublet which is well separated in energy to other crystal field 
		levels.  
		This quasi-doublet accounts for the low-temperature magnetism of the 
		material.}
	
	\label{fig:CEF}
\end{figure}
\hide{
\begin{figure}
	\includegraphics[width=1\linewidth]{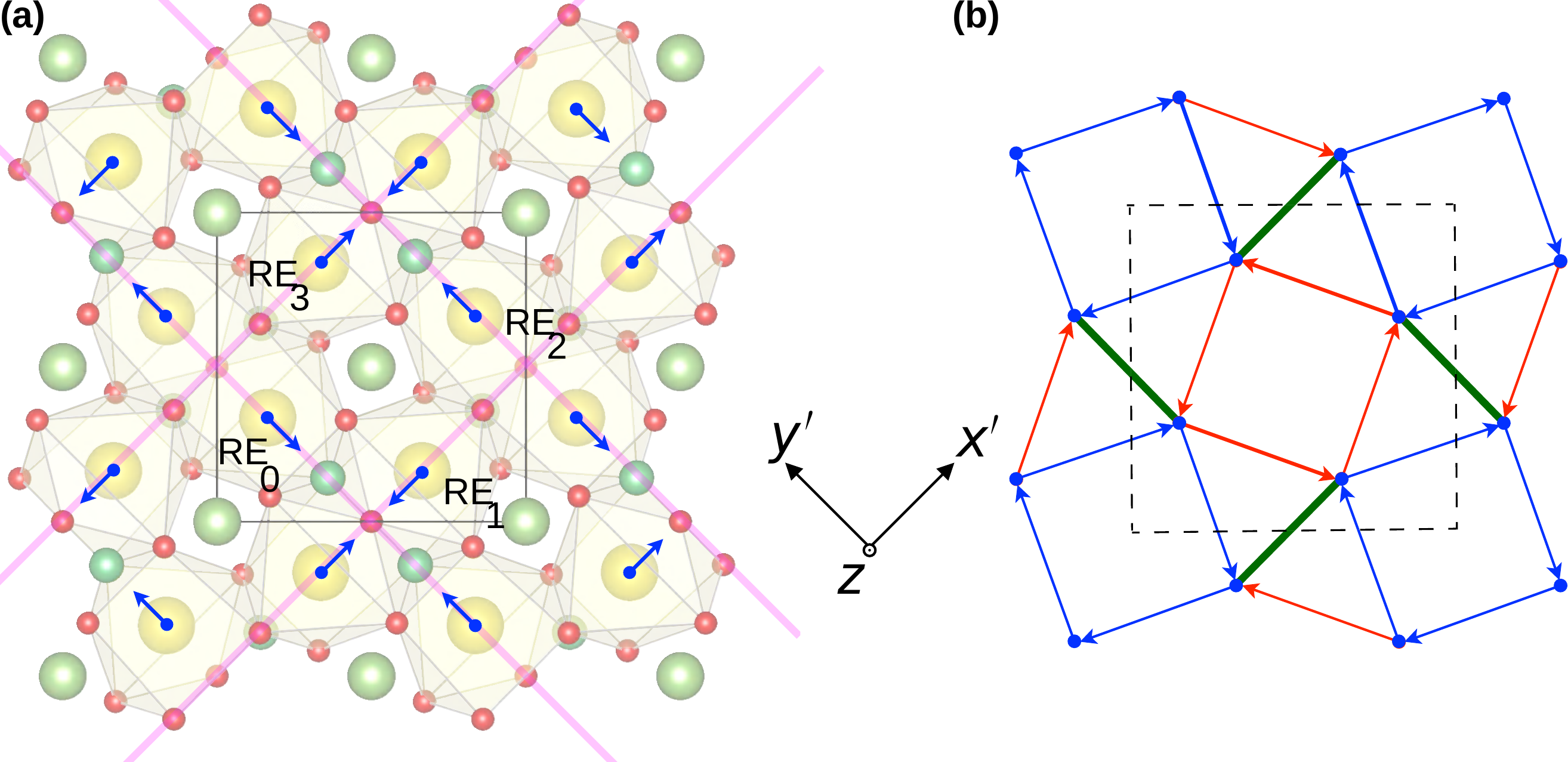}
	\caption{Crystal structure of RE$_{2}$Ga$_{2}$BeO$_{7}$ and sketch of 
	the SSL. (a). RE$^{3+}$ and O$^{2-}$ are denoted by yellow and red balls, respectively. Mirror
		planes of RE$^{3+}$ are indicated by purple lines. The in-plane 
		direction of the local dipole axes $\mathbf{n}_{i}$ are labeled by blue arrows. 
		(b). The intra-dimer bonds of the SSL are indicated by thick dark green 
		lines. The $\eta_{ij}=\pm1$ inter-dimer bonds in Eq. (\ref{eq:inter-dimer}) are marked as red and blue, respectively, with the directions $i\rightarrow j$ are 
		marked by arrows.}
	\label{fig:crystal}
\end{figure}
}
\subsection{Local moment structures}

The local moment structures of non-Kramers ions can be analyzed from
symmetries of the crystal field levels. We define the $z$ diretion to be 
perpendicular to the plane of local moments, and $x^\prime$ and $y^\prime$ are 
along the two orthogonal dimer directions (see Fig. \ref{fig:crystal}(a)). Note 
that there are four RE$^{3+}$
ions within a unit cell (labeled as RE$_{0}$, RE$_{1}$, RE$_{2}$,
RE$_{3}$, respectively in Fig. \ref{fig:crystal}(a)) that are related by the 
four-fold roto-inversion $S_{4}$ about the $z$ axis. Without loss of generality,
we first focus on the RE$_{1}$ sublattice, and then expand our
results to the other three. As mentioned above, the point group symmetry
of each RE$^{3+}$ ion only contains a mirror reflection $\sigma_{v}$
with the mirror plane parallel to the corresponding dimer direction,
see Fig. \ref{fig:crystal}. The CEF calculations \cite{li2024spinons}
suggest that the quasi-doublet states $|\psi_{\pm}\rangle$ carry $A_{1}$
and $A_{2}$ representations of $\sigma_{v}$, respectively:
\[
\sigma_{v}:|\psi_{\pm}\rangle\rightarrow\pm|\psi_{\pm}\rangle.
\]
Here we omit the site indices for simplicity. We assume that the magnetism of 
the system can be understood within this low-energy sector. This two-level 
subsystem is described by an effective spin-1/2 model. The effective spin-1/2
operators can be constructed as Pauli matrices acting on the two-level
system: $\hat{\sigma}^{\alpha}\equiv 
\frac{1}{2}\sum_{\mu,\nu=\pm}|\psi_{\mu}\rangle\tau_{\mu\nu}^{\alpha}\langle\psi_{\nu}|
 =\frac{1}{2}\psi\tau^{\alpha}\psi^{\dagger}$ 
where $\psi=\left(|\psi_{+}\rangle,|\psi_{-}\rangle\right)^{T}$ and 
$\tau^\alpha$ $(\alpha=1,2,3)$ are Pauli matrices. 
Denote $\mathcal{P}\equiv \psi\psi^{\dagger} 
=|\psi_{+}\rangle\langle\psi_{+}|+|\psi_{-}\rangle\langle\psi_{-}|$ 
as the projection operator onto the two-level subspace. For each spin
component, the symmetry transformations under the mirror reflection
$\sigma_{v}$ and time-reversal $\Theta$ are as follows:

\[
\sigma_{v}:\hat{\sigma}^{1}\rightarrow-\hat{\sigma}^{1},\, 
\hat{\sigma}^{2}\rightarrow-\hat{\sigma}^{2},\,\hat{\sigma}^{3}\rightarrow\hat{\sigma}^{3},
\]

\[
\Theta:\hat{\sigma}^{1}\rightarrow-\hat{\sigma}^{1},\, 
\hat{\sigma}^{2}\rightarrow\hat{\sigma}^{2},\,\hat{\sigma}^{3}\rightarrow\hat{\sigma}^{3}.
\]
One sees that both $\hat{\sigma}^{2}$ and $\hat{\sigma}^{3}$ components
are even under time-reversal and therefore behave as electric quadrupoles.
Meanwhile, $\hat{\sigma}^{1}$ is odd under both time-reversal and 
the mirror reflection and corresponds to a 
magnetic dipole within the mirror plane (a pseudovector transforming
as $\mathcal{J}^{x^\prime}$ or $\mathcal{J}^{z}$ under crystalline symmetries). 
This result can be obtained in a more formal way by projecting the total 
angular momentum $\boldsymbol{\mathcal{J}}$ onto the two-level
subspace $\hat{\mathbf{j}}=\mathcal{P}\boldsymbol{\mathcal{J}}\mathcal{P}$ and 
rewrite the 
dipolar operator as $\hat{\mathbf{j}}=A\hat{\sigma}^{1}\mathbf{n}$,
where $A$ describes the magnitude of the dipole moment, and $\mathbf{n}$
is the unit vector representing the direction of the dipole axis.
Note that 
$\mathcal{J}^{x^{\prime}}$
and $\mathcal{J}^{z}$ transform according to the
same representation of the crystalline symmetry. Consequently,
the dipole 
direction $\mathbf{n}$ cannot be 
completely determined by symmetry,
but depends on microscopic details of crystal field wave functions.

The above discussions are restricted to the RE$_{1}$ ion. As the
four RE ions within the unit cell are connected by the four-fold roto-inversion
about $z$ axis, their individual local dipole axes $\mathbf{n}_{i}$
should be connected by the four-fold counter-rotation. Therefore,
the local dipole axes $\mathbf{n}_{i}$ at sublattice $i$ should have
identical out-of-plane component, while their in-plane components
follow the directions as shown in Fig. \ref{fig:crystal}. Note that
such sublattice-dependent structure of dipole moments makes their coupling
to external magnetic field peculiar, and complicates the analysis
of magnetization and magnetic susceptibility data, as will be discussed
in detail in Sec. \ref{sec:BField}.

\subsection{Effective Hamiltonian}

In this subsection, we derive a generic Hamiltonian based on symmetry
analysis. Considering the strongly localized nature of 4$f$ electronic
orbitals, we expect the exchange interactions decay rapidly with inter-moment 
distance. To keep the analysis as concise as possible, here we only give the 
results for the intra-dimer and nearest-neighbor (NN) inter-dimer and 
interactions. The interaction between further neighboring spins can be derived 
in a similar way. 

We start from derivation of the intra-dimer interactions. The dimer is protected
by two mirror reflection symmetries $\sigma_{v}$ and $\sigma_{v}^{\prime}$,
with the mirror planes parallel and perpendicular to the dimer directions,
respectively (see Fig.~\ref{fig:crystal}(a)). 
The most general form of the intra-dimer interactions 
can be written as 
bilinears of effective spin operators: 
$H_{J^{\prime}}=\sum_{\langle\langle 
ij\rangle\rangle\alpha\beta} J_{ij}^{\prime\alpha\beta} 
\hat{\sigma}_{i}^{\alpha}\hat{\sigma}_{j}^{\beta}$, where 
$J_{ij}^{\prime\alpha\beta}$ refers to the intra-dimer exchange coupling 
between $\alpha$ component of the effective spin located at sites $i$ and the 
$\beta$ component of the spin at site $j$.
Time-reversal symmetry forbids any mixing between dipole and quadrupole
moments. This immediately leads to 
$J_{ij}^{\prime12}=J_{ij}^{\prime13}=J_{ij}^{\prime21}=J_{ij}^{\prime31}=0$.
Moreover, the mirror reflection $\sigma_{v}$ forbids linear mixing of 
$\hat{\sigma}_{i}^{2}$
and $\hat{\sigma}_{j}^{3}$ components since they have different
parities. As a result, 
$J_{ij}^{\prime23}=J_{ij}^{\prime32}=0$.
Therefore, the intra-dimer interaction simply takes the form of an XYZ
model 
\begin{equation}
H_{J^{\prime}}=\sum_{\langle\langle 
ij\rangle\rangle}J^{\prime11}\hat{\sigma}_{i}^{1}\hat{\sigma}_{j}^{1} 
+J^{\prime22}\hat{\sigma}_{i}^{2}\hat{\sigma}_{j}^{2} 
+J^{\prime33}\hat{\sigma}_{i}^{3}\hat{\sigma}_{j}^{3}.
\end{equation}
Note that although the system does not contain any spatial-inversion
symmetries at the center of the dimer, the intra-dimer antisymmetric 
Dzyaloshinskii-Moriya 
interaction (DMI) \cite{DM1958,DM1960} is still disallowed by the spatial and time-reversal
symmetries. 

Next we consider the NN inter-dimer interaction. The interaction takes
the generic bilinear form $H_{J}=\sum_{\langle ij\rangle\alpha\beta}J_{ij}^{\alpha\beta}\hat{\sigma}_{i}^{\alpha}\hat{\sigma}_{j}^{\beta}.$
Time-reversal symmetry also forbids linear mixing between dipole and
quadrupole moments: $J_{ij}^{12}=J{}_{ij}^{13}=J{}_{ij}^{21}=J{}_{ij}^{31}=0$.
Since there is no other symmetry to further constrain the interactions,
the NN interaction takes a more complicated bond-dependent form
\begin{align}
H_{J} & =\sum_{\langle ij\rangle}\left(\hat{\sigma}_{i}^{1},\hat{\sigma}_{i}^{2},\hat{\sigma}_{i}^{3}\right)\begin{pmatrix}J^{11}\\
 & J^{22} & \eta_{ij}J^{23}\\
 & \eta_{ij}J^{32} & J^{33}
\end{pmatrix}\begin{pmatrix}\hat{\sigma}_{j}^{1}\\
\hat{\sigma}_{j}^{2}\\
\hat{\sigma}_{j}^{3}
\end{pmatrix}\label{eq:inter-dimer}
\end{align}
where $\langle ij\rangle$ follows the bond direction $i\rightarrow j$
as shown in Fig. \ref{fig:crystal}(b). The bond dependence of the inter-dimer
interactions is taken over by the off-diagonal $\eta_{ij}$ term,
which takes the values $1$ and $-1$ for the red and blue bonds, respectively.

Besides the exchange couplings, we have to consider the effect of crystal 
field, which causes a splitting $\Delta$ between the two levels of the 
quasi-doublet. This can be taken as  
an intrinsic field acting on the quadrupole component 
$\hat{\sigma}^{3}$ \cite{liu2020intrinsic,shen2019intertwined}
\begin{equation}
H_{\Delta}=-\Delta\sum_{i}\hat{\sigma}_{i}^{3}.
\end{equation}
The total Hamiltonian then takes the following form
\begin{equation}
H=H_{J^{\prime}}+H_{J}+H_{\Delta}.\label{eq:ham}
\end{equation}

\hide{
\begin{figure}
	\includegraphics[width=0.73\columnwidth]{bond}
	
	\caption{Illustration of the Shastry-Sutherland lattice. The intra-dimer bonds are indicated by thick dark green lines. The $\eta_{ij}=\pm1$
		inter-dimer bonds Eq. (\ref{eq:inter-dimer}) are marked are marked
		red and blue, with the directions $i\rightarrow j$ are marked by
		arrows.}
	
	\label{fig:bond}
	\end{figure}
}

\begin{table*}[t]
	\begin{ruledtabular}
		\begin{tabular}{cccc}
			Field direction & Induced magnetization & $C$ & 
			$\Theta_{CW}$\tabularnewline
			\colrule{[}001{]} & 
			$\mu_{B}g_{J}A_{\perp}\left(\hat{\sigma}_{0}^{1}+\hat{\sigma}_{1}^{1}+\hat{\sigma}_{2}^{1}+\hat{\sigma}_{3}^{1}\right)$
			 & $k_{B}^{-1}(\mu_{B}g_{J}A_{\perp})^{2}/4$ & 
			$-k_{B}^{-1}(4J^{11}+J{}^{\prime11})/4$\tabularnewline
			{[}100{]} & 
			$\mu_{B}g_{J}A_{\parallel}/\sqrt{2}\left(\hat{\sigma}_{0}^{1}-\hat{\sigma}_{1}^{1}-\hat{\sigma}_{2}^{1}+\hat{\sigma}_{3}^{1}\right)$
			 & $k_{B}^{-1}(\mu_{B}g_{J}A_{\parallel})^{2}/8$ & 
			$k_{B}^{-1}J{}^{\prime11}/4$\tabularnewline
			{[}010{]} & 
			$\mu_{B}g_{J}A_{\parallel}/\sqrt{2}\left(-\hat{\sigma}_{0}^{1}-\hat{\sigma}_{1}^{1}+\hat{\sigma}_{2}^{1}+\hat{\sigma}_{3}^{1}\right)$
			 & $k_{B}^{-1}(\mu_{B}g_{J}A_{\parallel})^{2}/8$ & 
			$k_{B}^{-1}J{}^{\prime11}/4$\tabularnewline
			{[}110{]} & 
			$\mu_{B}g_{J}A_{\parallel}\left(-\hat{\sigma}_{1}^{1}+\hat{\sigma}_{3}^{1}\right)$
			 & $k_{B}^{-1}(\mu_{B}g_{J}A_{\parallel})^{2}/8$ & 
			$k_{B}^{-1}J{}^{\prime11}/4$\tabularnewline
			{[}$\bar{1}$10{]} & 
			$\mu_{B}g_{J}A_{\parallel}\left(-\hat{\sigma}_{0}^{1}+\hat{\sigma}_{2}^{1}\right)$
			 & $k_{B}^{-1}(\mu_{B}g_{J}A_{\parallel})^{2}/8$ & 
			$k_{B}^{-1}J{}^{\prime11}/4$\tabularnewline
		\end{tabular}
	\end{ruledtabular}

	\caption{Induced magnetization, Curie-Weiss parameter $C$ and Curie-Weiss
		temperature $\Theta_{CW}$ for different field directions. Here 
		$A_{\perp}\equiv 
		A\left(\mathbf{n}_{i}\cdot\hat{\mathbf{e}}_{\perp}\right)$
		and $A_{\parallel}\equiv 
		A\left(\mathbf{n}_{i}\cdot\hat{\mathbf{e}}_{\parallel}\right)$
		denote the magnitudes of out-of-plane and in-plane dipole components
		$\hat{\sigma}^{1}$, respectively.}
	
	\label{tab:magnetization}
\end{table*}
 
\section{Coupling to the magnetic field\label{sec:BField}}

In this section, we discuss the implications of the sublattice-dependent
dipole moment structure when coupled to an external magnetic field.
Denoting the sublattice index as $i$ ($i=0,1,2,3$ for the four sites within a 
unit cell), the general form of Zeeman
coupling takes the form
\begin{align}
H_{B} & =-\mu_{B}g_{J}\sum_{\alpha}B^{\alpha}\sum_{i=0}^{3}\hat{j}_{i}^{\alpha}\nonumber \\
 & =-\mu_{B}g_{J}\sum_{\alpha}B^{\alpha}\sum_{i=0}^{3}\left(\mathbf{n}_{i}\cdot\hat{\mathbf{e}}_{\alpha}\right)\hat{j}_{i}^{\mathbf{n}_{i}}\nonumber \\
 & =-\mu_{B}g_{J}\sum_{\alpha}B^{\alpha}\sum_{i=0}^{3}\left(\mathbf{n}_{i}\cdot\hat{\mathbf{e}}_{\alpha}\right)A\hat{\sigma}_{i}^{1}.
\end{align}
Here $B^{\alpha}$ denotes the applied magnetic field along the $\alpha$ 
direction, $\alpha=x,y,z$ corresponds
to the spatial directions where $x,y$ are obtained from $x^\prime,y^\prime$ by 
a rotation of $45^\circ$ about the $z$ axis. $g_{J}$ is the Land\'e g-factor, 
$\hat{j}_{i}^{\alpha}$ is the 
projection of
the total angular momentum $\mathcal{J}_{i}^{\alpha}$ onto the low-energy 
subspace:
\begin{equation}
\hat{j}_{i}^{\alpha}\equiv\mathcal{P}_{i}\mathcal{J}_{i}^{\alpha}\mathcal{P}_{i}.
\end{equation}
Given that the effective spin components $\hat{\sigma}_{i}^{2}$ and 
$\hat{\sigma}_{i}^{3}$ are even under time-reversal, only the dipolar
component $\hat{\sigma}_{i}^{1}$ directly couples to the external
magnetic field. More interestingly, because the dipole axis direction 
$\mathbf{n}_{i}$
exhibits sublattice dependence, the in-plane and out-of-plane field couples
to $\hat{\sigma}_{i}^{1}$ differently. The out-of-plane magnetic
field couples to the uniform magnetization of the dipolar spin component
$\hat{\sigma}_{0}^{1}+\hat{\sigma}_{1}^{1}+\hat{\sigma}_{2}^{1}+\hat{\sigma}_{3}^{1}$
just like in ordinary magnets, but the in-plane magnetic field turns out
to couple to the staggered magnetization of the dipolar moments. For example, 
the in-plane field $\mathbf{B}\parallel{[100]}$
couples to $\hat{\sigma}_{0}^{1} -\hat{\sigma}_{1}^{1} 
-\hat{\sigma}_{2}^{1}+\hat{\sigma}_{3}^{1}$. In Tab. \ref{tab:magnetization} we 
have listed several more examples on how the field couples to the dipolar 
moments.

This peculiar coupling of dipole moments to the external magnetic
field makes the behavior of magnetic susceptibilities quite different
from conventional magnetic materials. At high-temperatures, the magnetic
susceptibility per RE site should satisfy the Curie-Weiss (CW) law 
\begin{equation}
\chi\approx\frac{C}{T-\Theta_{CW}},
\end{equation}
where $\Theta_{CW}$ is the CW temperature and $C$ is the CW parameter. In 
Appendix \ref{sec:Derivation-of-Curie-Weiss}, we have derived the
CW behaviors for different field directions and the results
are summarized in Tab. \ref{tab:magnetization}. As the in-plane and out-of-plane
fields couple to different magnetization patterns, the corresponding
CW temperatures turns out to reflect interactions along different
bonds. This behavior is quite different from ordinary magnets
where the in-plane and out-of-plane CW temperatures reflect
the magnetic exchanges along corresponding directions. The above results imply 
that we could extract the
values of $J^{11}$ and $J^{\prime11}$ exchange interactions from the 
experimental susceptibility data. However, the exchange couplings associated 
with other effective spin components are not avialble because they reflect 
interactions among quadrupolar moments which do not directly couple to the 
external magnetic field.  

\section{Phase diagram\label{sec:Phase diagram}}

Before discussing the numerical phase diagram of the model Eq. (\ref{eq:ham}),
we first make a comparison of the generic effective models between
the Kramers \cite{Duan2024} and non-Kramers systems. The interactions between the
non-Kramers local moments are described by an extended XYZ model along
both intra- and inter-dimer bonds, which are very much like the case
of the Kramers doublets. Therefore, our following discussions on the
physical effects of exchange anisotropy will still be applicable to
Kramers systems. Meanwhile, there exists substantial differences of
non-Kramers systems from the Kramers counterpart, which are reflected
in the following two aspects.

First, the physical nature of the local moments differs significantly
between Kramers and non-Kramers systems. In Kramers systems, all effective
spin components are magnetic dipoles that are linearly coupled to the external 
magnetic field and their ordering is detectable
by neutron diffraction. In contrast,
for non-Kramers systems only one component of the effective spin behaves as
magnetic dipole, while the other two are quadrupoles that are invisible
to neutrons. If the quadrupole components exhibit some non-vanishing
order $\langle\hat{\sigma}^{2}\rangle\neq0$ or $\langle\hat{\sigma}^{3}\rangle\neq0$,
such quadrupole ordering would not be quite visible in conventional
measurements. This poses an interesting challenge on how to identify these 
hidden
quadrupole ordered phases in experiments. Another difference is that
in Kramers systems the crystal field levels exhibit Kramers degeneracy
protected by the time reversal symmetry. Meanwhile, for non-Kramers systems 
there
is no such symmetry protection, hence the lowest two crystal field
levels have a finite splitting. This splitting is taken into account by an 
intrinsic field $H_{\Delta}=-\Delta\sum_{i}\hat{\sigma}_{i}^{3}$ that acts
on the quadrupole component $\hat{\sigma}^{3}$. In real
materials the magnitude of crystal field splitting are usually non-negligible
compared to the exchange interactions, therefore this intrinsic
field term must be seriously taken into account.

\subsection{Anisotropic XXZ and XYZ model with zero intrinsic field $\Delta=0$}

Previous numerical studies on 
the SSL magnets are mainly
focused on the isotropic Heisenberg model. Given the low symmetry
of systems we considered in this work, the effective Hamiltonian Eq. 
(\ref{eq:ham}) exhibits
highly anisotropic interactions with a large number of free parameters.
To make our analysis more manageable, we first ignore the intrinsic
field $\Delta=0$ and begin by exploring the phase diagram in the presence
of XXZ anisotropy. Then we investigate the effects of the XYZ anisotropy
to the system. Note that the model Eq. (\ref{eq:ham}) at zero intrinsic
field limit $\Delta=0$ is quite analogous to that of the Kramers
system, despite the different physical nature of the effective spin
components. So our discussion here is also applicable to the Kramers
system. The role of non-zero intrinsic field will be discussed in
the next subsection.

For the XXZ model, to simplify our discussions, we assume that the
exchange anisotropy $\delta$ are the same for the intra- and inter-dimer
interactions:
\begin{align}
H & =\sum_{\langle\langle ij\rangle\rangle} 
J^{\prime}(\text{\ensuremath{\delta}}\hat{\sigma}_{i}^{1}\hat{\sigma}_{j}^{1} 
+\hat{\sigma}_{i}^{2}\hat{\sigma}_{j}^{2} 
+\hat{\sigma}_{i}^{3}\hat{\sigma}_{j}^{3})\nonumber \\
 & +\sum_{\langle ij\rangle}J(\delta\hat{\sigma}_{i}^{1}\hat{\sigma}_{j}^{1} 
 +\hat{\sigma}_{i}^{2}\hat{\sigma}_{j}^{2} 
 +\hat{\sigma}_{i}^{3}\hat{\sigma}_{j}^{3}),\label{eq:XXZ}
\end{align}
where the anisotropy parameter $\delta$ interpolates between the
XY ($\delta=0$) and Ising limit ($\delta\rightarrow\infty$). This model has a 
global U(1)
symmetry generated by the global effective-spin rotation of arbitrary
angle $\theta$ respect to the $\hat{\sigma}^{2}$-$\hat{\sigma}^{3}$
plane. To calculate the phase diagram, we employ the variational optimization
of the 16-PESS (Projected Entangled Simplex States) tensor network
states \cite{chen2020automatic,liao2019differentiable,ponsioen2022automatic}.
In this context, the 16-PESS structure of the SSL
incorporates $U$ and $S$ tensors \cite{xi2023plaquette}. The $U$
tensor is defined within the plaquettes containing intra-dimer interactions
and is connected by the entangled simplex tensor $S$. The $S$ tensor,
which does not carry any physical spin degrees of freedom, is introduced
to describe the entanglement among the spin clusters. We performed
the calculations within a $2\times2$ unit cell, and employed translational
symmetry throughout the process. We utilized the corner transfer matrix
renormalization group (CTMRG) \cite{corboz2014competing} for infinite
network contraction. All results via the variational optimization
are obtained by using a bond dimension $D=4$ and CTMRG truncation dimension
$\chi=50$.

\begin{figure}
\includegraphics[width=1\linewidth]{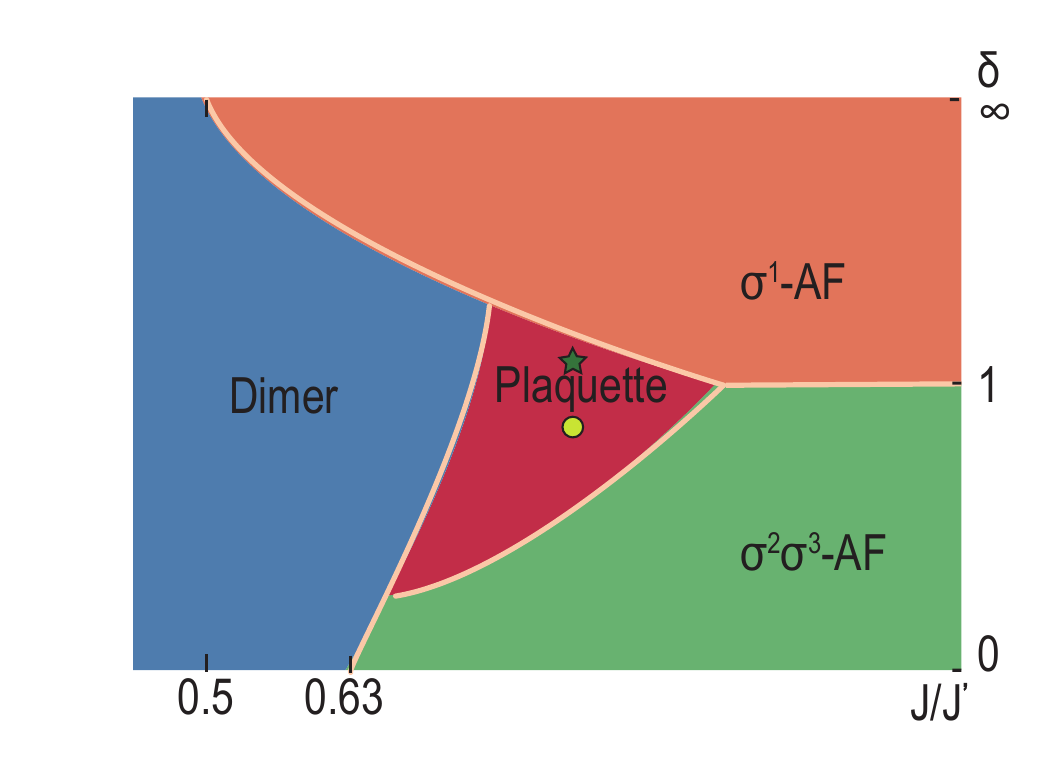}

\caption{Phase diagram of the antiferromagnetic XXZ model Eq. (\ref{eq:XXZ})
with zero intrinsic field $\Delta=0$. For non-Kramers systems $\hat{\sigma}^{1}$
is a dipole and $\hat{\sigma}^{2}$ and $\hat{\sigma}^{3}$ are quadrupoles,
$\sigma^{1}$-AF corresponds to an AFM state with ordering of the $\sigma^{1}$ 
component, and $\sigma^{2}\sigma^{3}$-AF corresponds to an AFQ state with 
ordering of the $\sigma^{2,3}$ components.}

\label{XXZ Phase diagram}
\end{figure}

The numerical phase diagram of the XXZ model Eq. (\ref{eq:XXZ}) is
presented in Fig. \ref{XXZ Phase diagram}. We find that the dimerized and
plaquette phases in the SSL Heisenberg model are
still stable against XXZ anisotropy. In the Ising anisotropic case ( 
$\delta>1$), the AFM phase in the Heisenberg case remains, which is now denoted 
as the $\sigma^1$-AF phase. But for the XY anisotropic case ($\delta<1$),  
the ordering of $\sigma^2$ or $\sigma^3$ (denoted as $\sigma^{2,3}$-AF phase) 
actually corresponds to an antiferro-quadrupolar (AFQ) order.
Specifically, when $J/J^{\prime}$
is relatively small, the dimer phase can remain stable throughout
the entire parameter space of $\delta$. In the $\delta\rightarrow0$ limit, we 
find the dimerized phase undergoes a direct transition to the 
$\sigma^{2,3}$-antiferro-quadrupole (AFQ) phase at $J/J^{\prime}=0.63$. 
In the Ising limit ($\delta\rightarrow\infty$) the transition between the 
dimerized and the $\sigma^1$-AF phases os at $J/J^\prime=0.5$, consistent with 
the value of previous theoretical studies 
\cite{kairys2020simulating,verkholyak2014exact}. 
The plaquette phase can remain stable
over a certain range of anisotropy before transitioning to the $\sigma^{1}$-AFM
or $\sigma^{2,3}$-AFQ phase. 

\begin{figure}
\includegraphics[width=1\linewidth]{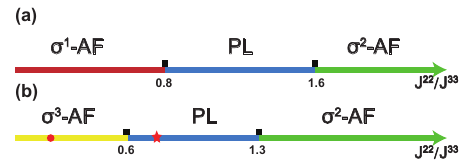}

\caption{Phase diagrams of the antiferromagnetic XYZ model with zero intrinsic
field $\Delta=0$. For non-Kramers systems, $\hat{\sigma}^{1}$-AF
is an AFM order, while $\hat{\sigma}^{2}$( $\hat{\sigma}^{3}$)-AF
are AFQ orders. (a) Phase diagram with $J^{22}/J^{33}$ at $J^{\prime22}=1$, 
$J^{11}=0.816,\ J^{33}=0.68,\ J^{\prime11}=1.2,\ J^{\prime33}=1.0$. The ground 
state for $J^{22}=J^{33}$ corresponds to the green star in Fig. \ref{XXZ Phase 
diagram}. (b) Phase diagram with with $J^{22}/J^{33}$ at $J'_{22}=1$, 
$J^{11}=0.544,\ J^{33}=0.68,\ J^{\prime11}=0.8,\ J^{\prime33}=1.0$. The ground 
state for $J^{22}=J^{33}$ corresponds to the yellow circle
in Fig. \ref{XXZ Phase diagram}.}

\label{XYZ Phase diagram}
\end{figure}

In systems corresponding to real materials, the exchange anisotropy is of an 
extended XYZ form that has even lower symmetry than the above XXZ model, with 
the global U(1) symmetry explicitly broken. As the generic Hamiltonian
Eq. (\ref{eq:ham}) at $\Delta=0$ still contains many free parameters,
we simplify our discussions by ignoring the off-diagonal interactions,
setting $J^{\alpha\beta}=0$ for $\alpha\neq\beta$. In addition,
we fix the relative proportions of the corresponding inter-dimer interactions
to match those of the intra-dimer ones, \emph{i.e.}, $J^{11}/J^{33}=J^{\prime11}/J^{\prime33}$
and $J^{22}/J^{33}=J^{\prime22}/J^{\prime33}$. Since the plaquette
phase has the strongest quantum fluctuations among the four phases of the phase 
diagram in Fig.~\ref{XXZ Phase diagram}, 
we select two representative parameter points within the plaquette phase in the 
XXZ
phase diagram (depicted by the green star and yellow circle in Fig.
\ref{XXZ Phase diagram}), and calculate the ground-state phase diagram in the 
presence of finite XYZ anisotropy $J^{22}\neq J^{33}$. From the calculated 
phase diagrams shown in Fig.~\ref{XYZ Phase diagram},
we find that because of its gapped nature, the plaquette phase remains stable 
when the XYZ anisotropy is not strong. As the anisotropy further increases, the
system ultimately enters to either AFM or AFQ phase depending on the 
anisotropic factors.

\subsection{Non-zero intrinsic field $\Delta\neq0$}

\begin{figure}[b]
	\includegraphics[width=1\linewidth]{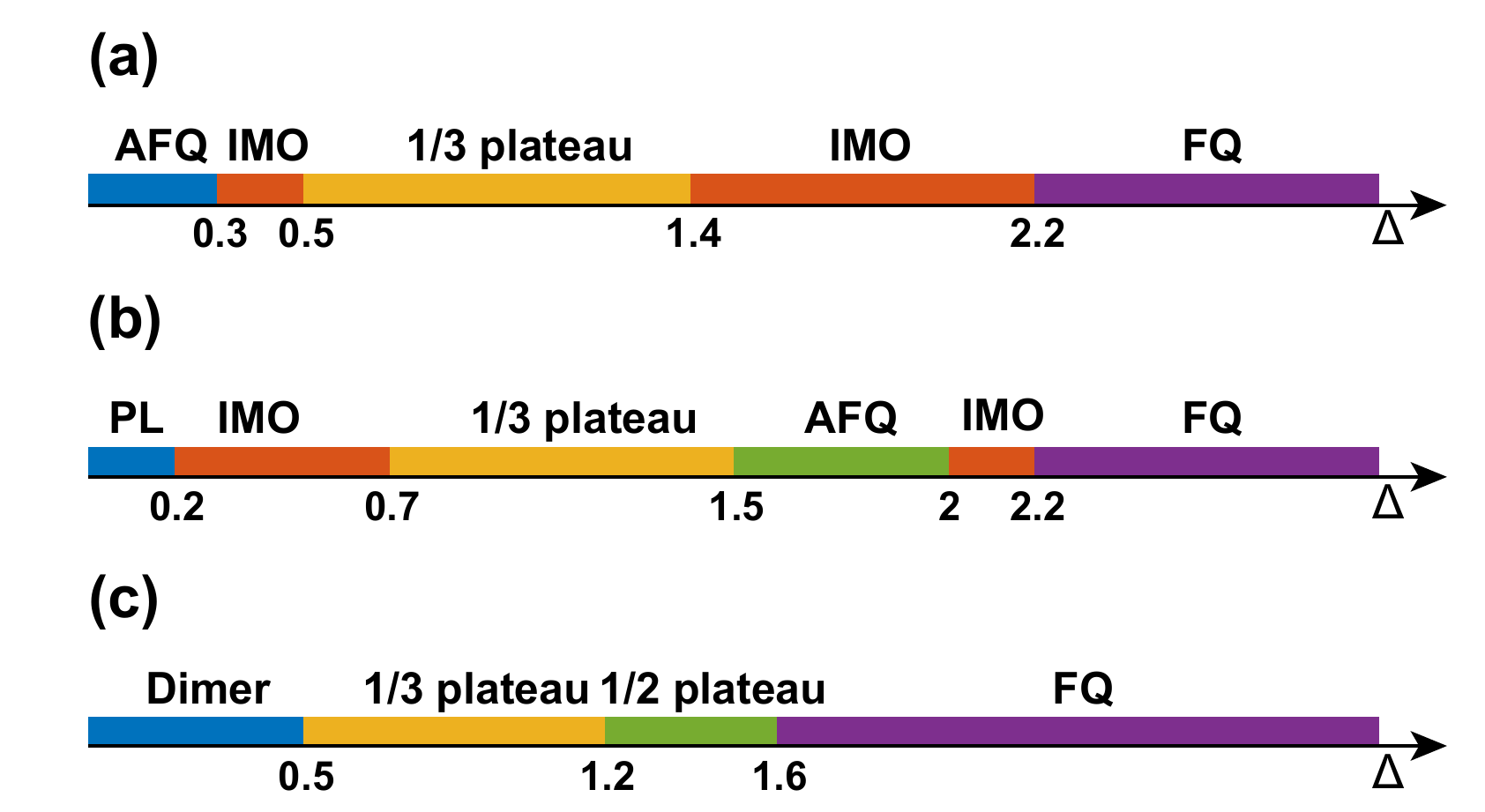}
	\caption{Phase diagrams of the non-Kramers model Eq. 
	(\ref{eq:ham}) versus
		the intrinsic field $\Delta$. The parameters are chosen such that
		the zero-field ground states are (a) the AFQ phase with 
		$J^{11}=0.544,\ J^{22}=0.136,\ J^{33}=0.68,\ J^{\prime11}=0.8,\ 
		J^{\prime22}=0.2,\ J^{\prime33}=1.0$;
		(b) the plaquette phase, with $J^{11}=0.544,\ J^{22}=0.51,\ 
		J^{33}=0.68,\ J^{\prime11}=0.8,\ J^{\prime22}=0.75,\ J^{\prime33}=1.0$;
		and (c) the dimerized phase with $J^{11}=0.4,\ J^{22}=0.3,\ 
		J^{33}=0.5,\ 
		J^{\prime11}=0.8,\ J^{\prime22}=0.6,\ J^{\prime33}=1.0$, respectively. 
		FQ, AFQ, and IMO refer to ferro-quadrupolar order, 
		antiferro-quadrupolar order, and 
		intertwined multipolar order, respectively.}

\label{fig:phasediagram_plateau}
\end{figure}

\begin{figure*}
	\includegraphics[width=1\linewidth]{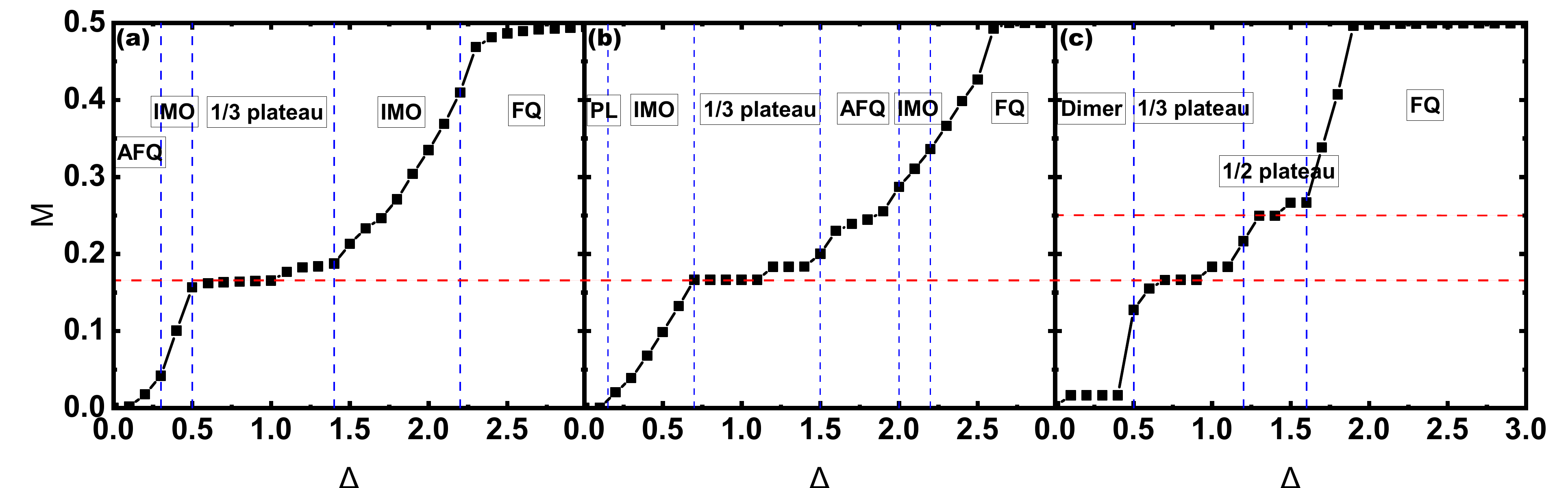}
	
	\caption{Quadrupole polarization $M$ versus the intrinsic field $\Delta$. 
	The parameters are chosen as same as those in Fig. 
		\ref{fig:phasediagram_plateau}. The dashed line indicates $M$ reaches 
		$1/3$ and $1/2$ of its saturated value.
		}
	
	\label{Magnetic Curve}
\end{figure*}

Here we focus on the non-Kramers system and investigate the fate of
phase diagram under a finite intrinsic field $\Delta\neq0$ that acts
on the quadrupole effective-spin component $\hat{\sigma}^{3}$. As
the intrinsic field is resulted from single-ion physics of crystal
field splitting, in general it should not be negligible compared with
the exchange interactions hence must be seriously taken into account.
To illustrate the physical effects of $\Delta$, we have selected
a few representative parameter points (the red circle and star in Fig.
\ref{XYZ Phase diagram}, for example), such that the zero-field limit $\Delta=0$
locate within the dimer, plaquette and AF phases, respectively. To
account for the possibility of incommensurate magnetic order when
an external field is applied, we perform the calculation by using the density 
matrix renormalization group (DMRG) algorithm 
\cite{stoudenmire2012studying,wang2023plaquette}
here. The calculation is performed on a $W\times L$ cylinder with
width $W$ and length $L$. Previous works on the Heisenberg model
\cite{lee2019signatures,wang2023plaquette} indicates that the minimal
width required to realize the plaquette singlet phase is $W=6$. Here,
we adopt $W=6$ and set the length $L=30$. The bond dimension
$D$ kept is equal to 1024. The calculated phase diagrams versus the
intrinsic field $\Delta$ are shown in Fig. \ref{fig:phasediagram_plateau},
and the corresponding quadrupolar polarization 
$M\equiv\frac{1}{N}\sum_{i}\hat{\sigma}^{3}$
versus $\Delta$ curves are shown in Fig. \ref{Magnetic Curve}.

The evolution of the isotropic Heisenberg system under an external magnetic
field has been extensively studied in the context of the SrCu$_{2}$(BO$_{3}$)$_{2}$
compound both experimentally \cite{kageyama1999exact,nomura2023unveiling, 
onizuka20001,takigawa200418,levy2008field,jaime2012magnetostriction}
and numerically \cite{muller2000exact,dorier2008theory,shi2022discovery, 
corboz2014crystals,huang2013dynamic}.
The magnetic field induces a rich phase diagram with many fractional 
magnetization plateau phases, such as 1/5, 1/4, 1/3, 1/2 plateaus, reflecting
the strong geometric frustration of the SSL.
Similarly, for the non-Kramers system with strong spin anisotropy we have also 
observed complicated evolution of ground states under
the intrinsic field $\Delta$.

\begin{figure}
	
	\includegraphics[width=1\linewidth]{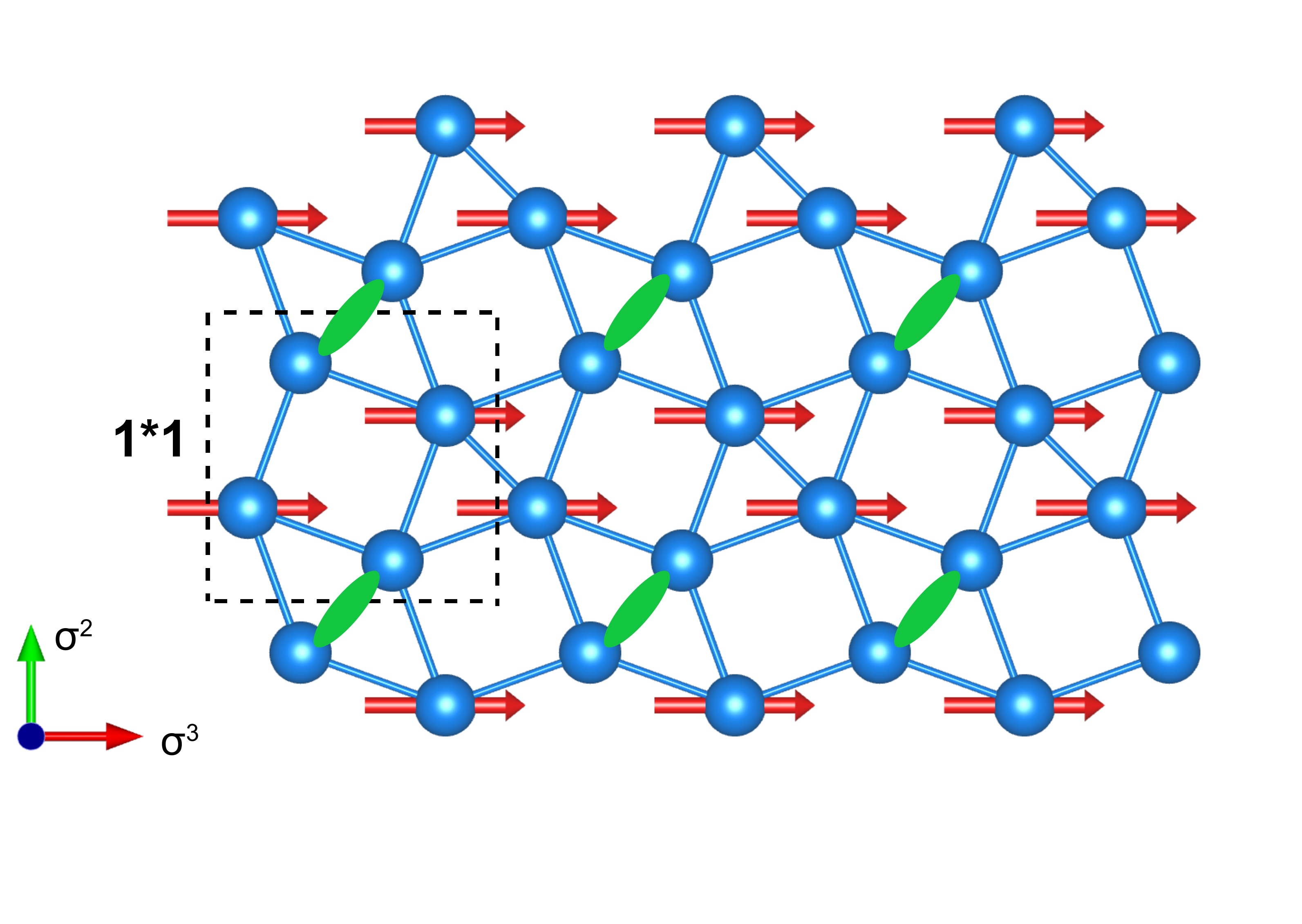}\caption{Effective-spin 
		configuration of the 1/2-\textquotedblleft plateau\textquotedblright{}
		phase. The green bond represents a singlet dimer and the red arrow 
		indicates the FQ order of effective spins induced by the intrinsic 
		field.}
	
	\label{1/2 plateau}
\end{figure}

For the plaquette and AFQ phases at zero $\Delta$, the field-induced
behaviors are qualitatively quite similar as shown in Fig. \ref{fig:phasediagram_plateau}(a)
and (b). As both phases are gapped, they are stable against small
fields $\Delta$. With increasing $\Delta$, the plaquette/AFQ phases
become unstable, developing coexisting AFQ and AFM orders which is denoted as
an ``intertwined multipolar order'' (IMO) 
\cite{shen2019intertwined,liu2018selective}.
Our DMRG results show that the ordering vectors of both dipole and quadrupole
components $\hat{\sigma}^{1}$ and $\hat{\sigma}^{3}$ are incommensurate. Then
it enters a peculiar type of $3\times1$ AFQ with approximate $1/3$
quadrupole polarization. This phase is the counterpart of 
the ``1/3-plateau'' phase  
of the Heisenberg system \cite{onizuka20001,shi2022discovery},
and its nature will be discussed in detail in the next subsection. 
After across an another IMO phase, the system eventually becomes completely 
polarized under sufficiently large $\Delta$. 
On the other hand, starting from the dimerized state,
the in-field phase diagrams is much simpler. The system subsequently
enters the ``1/3-plateau'' phase and a ``1/2-plateau'' phase, before fully 
polarized by the intrinsic field. In the ``1/2-plateau'' plateau phase, the spin
configuration is a defected FQ state. The magnitude of the magnetic
moments exhibits a distribution with a $(\pi,\pi)$ periodicity. The real-space 
pattern of effective spins of this state is shown in Fig. \ref{1/2 plateau}: In
one unit cell, one dimer forms a spin singlet state, while the effective spins 
of the other dimer are polarized by the intrinsic field and form an FQ order.

\subsection{Hidden \textquotedblleft 1/3-plateau\textquotedblright{} phase}

\begin{figure}
	\includegraphics[width=1\columnwidth]{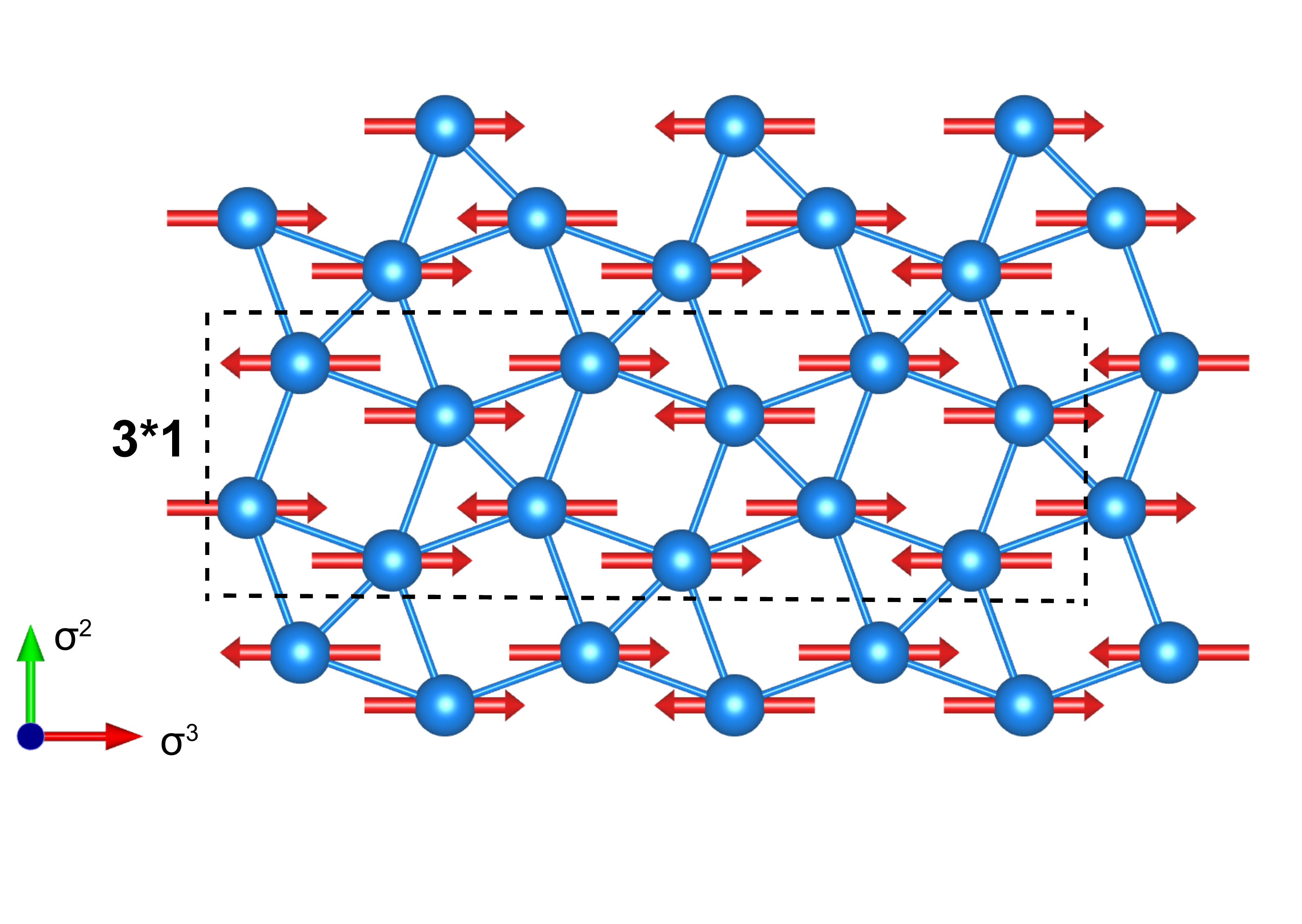}
	
	\caption{Effective-spin configuration of the 1/3-\textquotedblleft 
	plateau\textquotedblright{}
		phase.}
	
	\label{fig:spinconf_plateau}
\end{figure}

In the phase diagrams of Fig.~\ref{fig:phasediagram_plateau} we observe wide 
existence of the ``1/3-plateau''
phase stabilized by the intrinsic field $\Delta$. Its effective spin
structure is identical to that of the Heisenberg systems: the spins
form a $3\times1$ superstructure (see Fig. \ref{fig:spinconf_plateau}) where 
every third dimer forms a
triplet that aligns along the field direction while in the remaining
dimers two spins are alternatively aligned and anti-aligned along
the field. This state has six-fold degeneracy that breaks both the 
$\mathbb{Z}_{3}$ translational and $\mathbb{C}_{2}$ rotational symmetries. In 
the specific heat measurements, such a symmetry breaking could be reflected as 
some divergence behaviors upon varying temperatures. However, notice that such 
ordering occurs purely in the quadrupolar sector, hence is not directly visible 
from the conventional experimental probes such as neutron diffraction, which 
corresponds to one type of hidden order. 
Moreover, the intrinsic field comes from the crystal field splitting
which can be tuned by either applying external pressure or chemical doping.
But in either case the tunability is very limited. 
These difficulties pose an important question on how to identify this hidden 
quadrupolar ordered phase in experiments.

As proposed in Refs. \cite{liu2018selective,li2016hidden}, the information
of quadrupole orderings can be indirectly reflected in dynamical measurements
such as inelastic neutron scattering. This is due to the non-commutative
relation between the dipolar and the quadrupolar components. As neutron
spins only couple to the magnetic dipolar component $\hat{\sigma}^{1}$
at the linear order, in inelastic neutron scattering what is measured
is the dipole-dipole correlation
\begin{equation}
S^{11}(\mathbf{q},\omega)=\frac{1}{2\pi N}\sum_{ij}\int_{-\infty}^{+\infty} 
\textrm{d}t\,e^{i\mathbf{q}\cdot(\mathbf{r}_{i}-\mathbf{r}_{j})-i\omega 
t}\langle\hat{\sigma}_{i}^{1}(0)\hat{\sigma}_{j}^{1}(t)\rangle.\label{eq:DSF}
\end{equation}
Fortunately, this dynamical dipole-dipole correlation is able to reflect the 
dynamics
of the $\langle\hat{\sigma}^{3}\rangle\neq0$ quadrupolar order. When
neutron spins measure the $\hat{\sigma}^{1}$ dipolar moment, they
induce spin flip events on the quadrupolar moments, thereby creating
coherent quadrupole-wave excitations. These excitations carry information
about the underlying quadrupolar orderings. Thus, although quadrupolar
ordering itself cannot be observed in experiments, their dynamical
excitations can be indirectly probed.

\begin{figure}
\includegraphics[width=1\columnwidth]{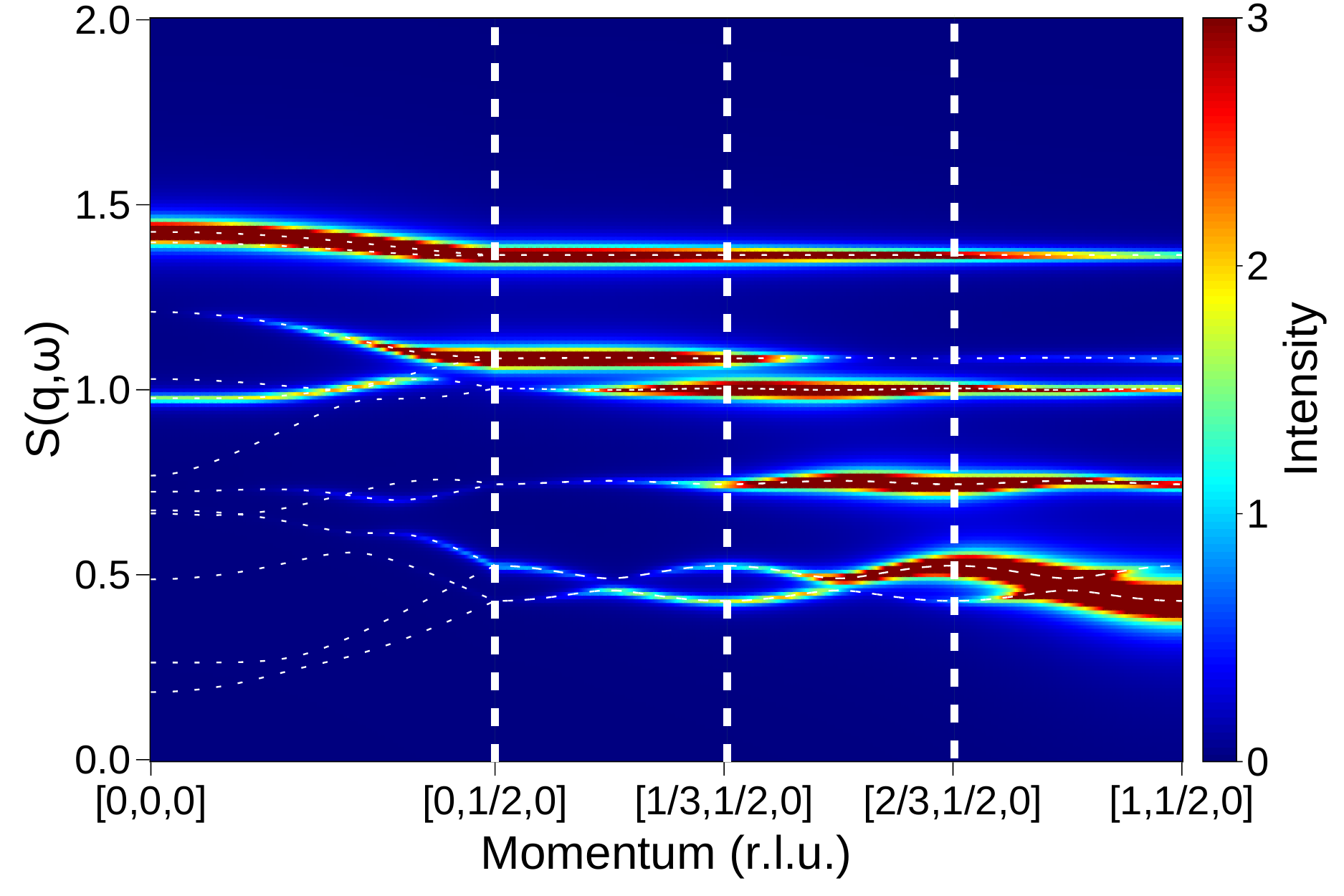}

\caption{Dynamical structural factor $S^{11}(\mathbf{q},\omega)$ for the 
``1/3-plateau''
phase. Spin-wave dispersions are marked by thin dashed lines. The model 
parameters are 
$J^{11}=0.544,\ J^{22}=0.136,\ J^{33}=0.68,\ J^{\prime11}=0.8,\ J^{\prime22}=0.2,\ J^{\prime33}=1.0,\ \Delta=0.9$.}

\label{fig:dynamics_plateau}
\end{figure}

Here we use the linear spin-wave theory to calculate the dynamical spin
structure factor $S^{11}(\mathbf{q},\omega)$ which can be directly
measured in inelastic neutron scattering experiments. The spin-wave
calculation is performed with the SUNNY package \cite{barros2022contributing},
and the excitation spectra are shown in Fig. \ref{fig:dynamics_plateau}.
In the spectra, we observe many branches of spin-wave dispersion,
consistent with the large magnetic supercell structure of the ``1/3-plateau''
phase. All spin-wave branches are rather flat, indicating that such
phase is rather classical with significantly suppressed quantum fluctuations.
This $3\times1$ structure is further reflected by the band 
folding
along the {[}0 1/2 0{]}-{[}1 1/2 0{]} line, indicating a period-of-three
structure along the {[}100{]} direction. Moreover, the excitations
become two-fold degenerate along the {[}0 1/2 0{]}-{[}1 1/2 0{]} line.
The reason of this degeneracy may be accounted by symmetries.

\section{Discussions and conclusions~\label{sec:Discussions}}

In this paper, we propose and study a generic spin model that describes
the interaction between the non-Kramers local moments relevant for
the Shastry-Sutherland lattice rare-earth magnets. We point out that the
local moments consist of both magnetic dipolar and quadrupolar components.
The effective model turns out to be an extended XYZ model with an intrinsic
field that accounts for the crystal field splitting. We discuss physical
effects of the XXZ and XYZ exchange anisotropy, as well as the intrinsic
field $\Delta$ to the system. Moreover, in the phase diagram we find
wide existence of pure quadrupolar orders that are invisible in conventional
experimental probes. We focus on the ``1/3-plateau'' pure quadrupolar
order and discuss its experimental signatures in dynamical excitations.

For the Shastry-Sutherland lattice, there are four spins per unit
cell, hence the Lieb-Shultz-Mattis theorem \cite{watanabe2015filling,hastings2004lieb,oshikawa2000comm,lieb1961two} do not apply to such systems.
Moreover, crystal field splitting can induce a finite intrinsic field
to the system. Therefore, absence of ordering signal in experiments
does not necessarily imply a non-trivial spin liquid ground state,
but can also be trivial quantum paramagnets such as dimer singlet
and the $\hat{\sigma}^{3}$-polarized state. Therefore, more experimental
probes such as Raman, X-ray scattering, neutron scattering and thermal
conductivity measurements, are needed to identify a possible spin
liquid state.

Here we discuss the relevance of our results with the non-Kramers
Shastry-Sutherland magnets Pr$_{2}$Be$_{2}$GeO$_{7}$ and 
Pr$_{2}$Ga$_{2}$BeO$_{7}$.
For Pr$_{2}$Be$_{2}$GeO$_{7}$, experiments find that the system
does not show any signatures of long-range magnetic ordering at temperatures
as low as 0.08 K, while the AC susceptibility reveal a dynamic spin-freezing
behavior below $\sim0.22$ K. These experimental signatures indicate
strong frustration of this system. However, from the Curie-Weiss temperatures
$\Theta_{\perp}=-0.33$ K and $\Theta_{\parallel}=-5.13$ K in the
experiment, we extract the exchange coupling $J^{11}\sim5.46$ K and
$J^{\prime11}\sim-20.52$ K, which is frustration-free in the dipole
sector $\hat{\sigma}^{1}$. Similar situations are also found in Pr$_{2}$Ga$_{2}$BeO$_{7}$.
Thermodynamics, $\mu$SR and neutron measurements indicate absence
of local moment ordering down to very low temperatures. Moreover,
a finite $\kappa/T$ is observed, suggesting the presence of
fermionic spinons at low energies. While these experimental signatures
also suggest strong frustration, the dipolar exchange couplings extracted
from the susceptibility of this material are found to be $J^{11}\sim41.3$
K and $J^{\prime11}\sim-85.4$ K, which also lacks frustration in the
dipole sector. 
\hide{To account for the experiments, it is essential to
	look into the dynamical frustration effects on the quadrupole sector
	in more detail, such as the XYZ anisotropy as well as the off-diagonal
	$\Gamma$ and/or Dzyaloshinskii-Moriya type quadrupole-quadrupole interactions
	($H_{\Gamma}\equiv\Gamma\sum_{\langle 
		ij\rangle}\eta_{ij}\left(\hat{\sigma}_{i}^{2}\hat{\sigma}_{j}^{3} 
	+\hat{\sigma}_{i}^{3}\hat{\sigma}_{j}^{2}\right)$
	and $H_{DM}\equiv D\sum_{\langle 
		ij\rangle}\eta_{ij}\left(\hat{\sigma}_{i}^{2}\hat{\sigma}_{j}^{3} 
	-\hat{\sigma}_{i}^{3}\hat{\sigma}_{j}^{2}\right)$)
	that are present in the effective model in Eq. (\ref{eq:ham}). }
	To account for the experiments, it is essential to
	look into the dynamical frustration effects on the quadrupole sector
	in more detail, such as the XYZ anisotropy and the inter-dimer
	$\Gamma$ and/or Dzyaloshinskii-Moriya type quadrupole-quadrupole interactions
	($H_{\Gamma}\equiv\Gamma\sum_{\langle 
		ij\rangle}\eta_{ij}\left(\hat{\sigma}_{i}^{2}\hat{\sigma}_{j}^{3} 
	+\hat{\sigma}_{i}^{3}\hat{\sigma}_{j}^{2}\right)$
	and $H_{DM}\equiv D\sum_{\langle 
		ij\rangle}\eta_{ij}\left(\hat{\sigma}_{i}^{2}\hat{\sigma}_{j}^{3} 
	-\hat{\sigma}_{i}^{3}\hat{\sigma}_{j}^{2}\right)$)
	that are present in the effective model in Eq. (\ref{eq:ham}). 
Including these terms further complicates the model, and we defer the study of 
the effects of them in a future work. 
Last but not least, more experimental probes, in particular dynamical 
spectroscopic measurements, are highly demanded to gain more insight into the 
nature of ground states of these systems.
 
\begin{acknowledgments}
	We thank Haidong Zhou, Xuefeng Sun and Jie Ma for useful discussions. 
	This work is supported by the National Key R$\&$D Program of China (Grant No.2023YFA1406500), and the National Science Foundation of China (Grant Nos.12334008 and 12174441). This paper is an outcome of ``Quantum Phase Transition in Low-Dimensional Frustrated Systems'' (RUC24QSDL038), funded by the ``Qiushi Academic - Dongliang'' Talent Cultivation Project at Renmin University of China in 2024.
\end{acknowledgments}

\appendix

\section{Curie-Weiss behaviors for non-Kramers systems\label{sec:Derivation-of-Curie-Weiss}}

In the following we analyze the high-temperature behavior of magnetic
susceptibilities in detail. For simplicity here we only consider four
RE sublattices within a unit cell, where the RE$_{i}$ site are labeled
by $i$ ($i=0,1,2,3$).

We first consider the out-of-plane magnetic field:
\begin{equation}
H_{[001]}=-\mu_{B}g_{J}A_{\perp}B^{[001]}\left(\hat{\sigma}_{0}^{1} 
+\hat{\sigma}_{1}^{1}+\hat{\sigma}_{2}^{1}+\hat{\sigma}_{3}^{1}\right).
\end{equation}
The out-of-plane magnetic field $B^{[001]}$ induces a uniform magnetization 
$m=\langle\hat{\sigma}_{i}^{1}\rangle$ 
at each site $i$. This magnetization can be regarded as the combined
effect of the external magnetic field and the internal exchange field:
\begin{equation}
m=\chi_{p}\left(\mu_{B}g_{J}A_{\perp}B^{[001]}+B_{exc}\right),
\end{equation}
where $\chi_{p}=(4k_{B}T)^{-1}$ is the Curie paramagnetic susceptibility,
$B_{exc}=-\left(4J^{11}+J^{\prime11}\right)m$ is the exchange field.
From the above relation we obtain
\begin{equation}
m=\frac{\chi_{p}\mu_{B}g_{J}A_{\perp}}{1+\chi_{p}(4J^{11}+J^{\prime11})}B^{[001]}.
\end{equation}
The high-temperature magnetic susceptibility per RE site satisfies the
Curie-Weiss relation
\begin{equation}
\chi^{[001]}\approx\frac{C}{T-\Theta_{CW}},
\end{equation}
with $C=k_{B}^{-1}(\mu_{B}g_{J}A_{\perp})^{2}/4$ and 
$\Theta_{CW}=-k_{B}^{-1}(4J^{11}+J^{\prime11})/4$.

Then we consider the effect of the in-plane magnetic field. We first consider
that the field is along the [100] direction.
\begin{equation}
H_{[100]}=-\mu_{B}g_{J}A_{\parallel}B^{[100]}/\sqrt{2}\left(\hat{\sigma}_{0}^{1}
 -\hat{\sigma}_{1}^{1}-\hat{\sigma}_{2}^{1}+\hat{\sigma}_{3}^{1}\right),
\end{equation}
which induces a staggered magnetization, say, $m$ on RE$_{1}$ and
RE$_{2}$ sublattices, and $-m$ on RE$_{0}$ and RE$_{3}$ ones.
The formation of the moment is due to the combined effect of the external
magnetic field and the internal exchange field:
\begin{equation}
m=\langle\hat{\sigma}_{1}^{1}\rangle 
=\chi_{p}\left(\mu_{B}g_{J}A_{\parallel}B^{[100]}/\sqrt{2}+B_{exc}\right),
\end{equation}
where the exchange field $B_{exc}=J{}^{\prime11}m$. From the above
relation we can obtain
\begin{equation}
m=\frac{\chi_{p}\mu_{B}g_{J}A_{\parallel}/\sqrt{2}} 
{1-\chi_{p}J{}^{\prime11}}B^{[100]}.
\end{equation}
The corresponding Curie-Weiss parameter and Curie-Weiss temperature
can then be obtained as $C=k_{B}^{-1}(\mu_{B}g_{J}A_{\parallel})^{2}/8$ and
$\Theta_{CW}=k_{B}^{-1}J^{\prime11}/4$.

At least we consider that the field is applied along the [110] direction.
The Zeeman coupling becomes:
\begin{equation}
H_{[110]}=-\mu_{B}g_{J}A_{\parallel}B^{[110]}\left(\hat{\sigma}_{3}^{1}-\hat{\sigma}_{1}^{1}\right).
\end{equation}
The applied magnetic field induces the staggered magnetization $m$ at
RE$_{3}$ site and $-m$ at RE$_{1}$ sites respectively, while it
does not couple to the moments of RE$_{0}$ and RE$_{2}$. The corresponding
Curie-Weiss parameters are $C=k_{B}^{-1}(\mu_{B}g_{J}A_{\parallel})^{2}/8$ and
$\Theta_{CW}=k_{B}^{-1}J^{\prime11}/4$, which are identical to those
for the {[}100{]} field.

\input{Theory_nonkramers1106.bbl}

\end{document}

%% file: Theory_nonkramers1106.bbl
%